\documentclass[twocolumn,prl,superscriptaddress,showpacs]{revtex4}
\begin{document}
\begin{abstract}
We present a perfect state transfer protocol via a qubit chain with the evolution governed by the $xx$ Hamiltonian. In contrast to the recent protocol announced in Phys. Rev. Lett. {\bf 101}, 230502 (2008), our method does not demand any remote-cooperated initialization and sending classical information about measurement outcomes. We achieve the perfect state transfer only with the assumption of access to two spins at each end of the chain, while the initial state of the whole chain is irrelevant. 
\end{abstract}
\title{Perfect State Transfer without State Initialization and Remote Collaboration}
\author{Marcin Markiewicz}\affiliation{Institute of Theoretical Physics and Astrophysics, University of Gda\'nsk, PL-80-952 Gda\'nsk, Poland} 
\author{Marcin Wie\'sniak}\affiliation{Institute of Theoretical Physics and Astrophysics, University of Gda\'nsk, PL-80-952 Gda\'nsk, Poland}
\date{12.05.2009}
\maketitle

The developing techniques of quantum computation \cite{DJ,Shor,Grover} are expected to accelerate some complex computational processes. However, when realized in solid-state systems such as Josephson junctions \cite{jj1,jj2,jj3}, or quantum dots \cite{qd1,qd2,qd3}, various modules of a quantum computer might need to faithfully exchange quantum information. This could be the case, e.g., when qubits are taken from the memory register to have gates applied on them. This anticipated requirement has stimulated interest in the problem of the quantum state transfer \cite{sugato}. It is a procedure, in which an arbitrary qubit state propagates over a lattice of spins-$\frac{1}{2}$ from one point to another (usually, between ends of a chain of $N$ spins-$\frac{1}{2}$). 

Initial efforts faced the problem of only quasi-periodicity of the free dynamics of the considered chains. This caused the quantum information to be transferred to be dislocated over the whole system. Hence the authors proposed protocols, which allowed to gather the full information about the quantum message by accessing only the end(s) of the chain(s) \cite{bgb,danielbose,danielbose2,poznan}. 

Christandl {\em et al.} \cite{christ} and independently Nikopolous {\em et al.} \cite{niko1,niko2} have found the Hamiltonian, which provides the perfect state transfer without any additional action from users. The Hamiltonian reads
\begin{eqnarray}
\label{Hxx}
H_{xx}&=&J\sum_{i=1}^{N-1}\sqrt{i(N-i)}(X^{[i]}X^{[i+1]}+Y^{[i]}Y^{[i+1]})\nonumber\\
&+&B\sum_{i=1}^{N}Z^{[i]},
\end{eqnarray}
where
$X=\frac{1}{2}\left(\begin{array}{cc} 0&1\\ 1&0\end{array}\right)$, $Y=\frac{1}{2}\left(\begin{array}{cc} 0&-i\\ i&0\end{array}\right)$, and $Z=\frac{1}{2}\left(\begin{array}{cc} 1&0\\ 0&-1\end{array}\right)$
are the spin-$\frac{1}{2}$ components in units where $\hbar=1$, square-bracketed superscripts label the qubit, $J$ is a constant, and $B$ is the magnetic field (to cancel a relative phase change). The evolution expressed by Hamiltonian (\ref{Hxx}) is periodic and such that if one of the spins at one side is pointing up ($|1\rangle$) rather than down (which is the initial direction of all other spins, $|0\rangle$), this {\em excitation} is found at the other side after half of the period. The key feature of this quantum wire, state mirroring, was further formalized by Shi {\em et al.} \cite{Shi} in the spectrum parity matching condition (SPMC). It states that in the one excitation subspace all even (with respect to the middle of the chain) energy eigenstates have energies expressed by even numbers (up to additive and multiplicative constants), while the odd states have energies related to odd numbers. This induces the relative phase flip between the odd and the even components at the half of the period. By initializing the wire in the fully magnetized state, and putting the first qubit in the transferred state, we obtain the same state at the receiver's site after some time.

More recently, the studies of the transfer have been done in the regime of limited access to a quantum wire. This is a natural assumption, as the system connecting the sender and the receiver is indeed a black box. For example, it is possible to reconstruct the coupling strengths just by observing the state of the first qubit \cite{danielkoji,paternostro2}. Another paper \cite{paternostro} demonstrates the transfer possibility without initializing the state of the interconnecting part of the wire. Such a result is particularly relevant if the coherence life time is long in comparison to the transfer time. The transfer should not be much affected by the external factors, but on the other hand, it would be impractical to use the same decoherence mechanisms to cool the wire down (effectively) to the pure, fully magnetized state. However, one should notice that the transfer described in \cite{paternostro} is still complex in terms of the communication cost. In this Brief Report we propose a similar transfer protocol, which is less demanding concerning the communication. In fact, each stage of our result is local and, neglecting the decoherence, does not require any measurement or conditional unitary transformation. This is in contrast to the method due to Di Franco {\em e. al.}.

Let us first briefly recall the result from \cite{paternostro}. For definiteness, but also for the sake of greater physical relevance, we only focus on the $xx$ interaction version. From the state mirroring property of  the system described by the Eq. (\ref{Hxx}) (let us hereafter take $B=0$ for simplicity) follows the evolution given below. In general, one has
\begin{eqnarray}
\label{ich15}
Z^{[i]}(t^*)=Z^{[N+1-i]}
\end{eqnarray}
and for $N$ even
\begin{eqnarray}
\label{ich2}
&X^{[i]}(t^*)X^{[N+1-i]}(t^*)=X^{[i]}X^{[N+1-i]},&\nonumber\\
&X^{[i]}(t^*)Y^{[N+1-i]}(t^*)=Y^{[i]}X^{[N+1-i]},&
\end{eqnarray}
while for $N$ odd one has 
\begin{eqnarray}
\label{ich3}
&X^{[i]}(t^*)X^{[N+1-i]}(t^*)=Y^{[i]}Y^{[N+1-i]},&\nonumber\\
&X^{[i]}(t^*)Y^{[N+1-i]}(t^*)=-X^{[i]}Y^{[N+1-i]}.&
\end{eqnarray}
The operators on the left-hand sides are given in the Heisenberg picture (time dependent) and $t^*=\frac{\pi}{J}$ is the transfer time. 

The sender announces the transfer and initializes his qubit in the state to be sent, while the receiver measures the $x$- (or $y$-) component of his spin (depending on the parity of $N$) obtaining the result $j=\pm \frac{1}{2}$. After the evolution, the states of the sender's and receiver's qubits are again factored out from the state of the rest of the wire. The sender measures the $x$-component of his spin and communicates the result $k=\pm \frac{1}{2}$. The receiver now applies a $jk$-dependent unitary transformation to the state of the last qubit and reconstructs the original state. One can now count the additional actions performed by the partners. The receiver conducted a measurement, which was coordinated in time with the encoding. Then, a measurement was done by the sender, who broadcasted the result. This result determined the receiver's choice of a unitary transformation. 

Interestingly, this scheme resembles the teleporation \cite{teleportation}, in which one qubit is sent at the expense of one measurement, two classical bits being transmitted and a conditional unitary transformation. Here, only one bit is sent and an entangled pair is not consumed. Instead, correlations produced during the transfer is destroyed. Since the initial state of the last qubit is $I\pm X^{[N]}$ $\left(I=\frac{1}{2}\left(\begin{array}{cc}1&0\\0&1\end{array}\right)\right)$, this entanglement arises from the evolution of operators $X^{[1]}, Y^{[1]}, Z^{[1]}X^{[N]},$ and $X^{[N]}$, which are not described by Eqns. (\ref{ich15}-\ref{ich3}). 

In what follows, we present our protocol for the state transfer without state initialization and remote collaboration. At the write-in stage, only the sender needs to operate on the wire; the state of the rest is irrelevant. This is achieved however, at the expense of access to two first and two last sites of the chain, as the proposed encoding is the two qubit codes. Similarly, as in the case of the dual-rail scheme \cite{danielbose}, the logical qubit space is spanned by the subspace of the null $z$-magnetization of the two qubits, $\langle Z^{[1]}+Z^{[2]}\rangle=0$. Logical $|0\rangle$ is represented by the first spin flipped upwards and the second flipped downwards, logical $|1\rangle$ corresponds to the opposite situation. Superpositions are translated into entangled states.

The operators, which encode the qubit according to the Bloch formula,
\begin{equation}
\label{bloch}
\rho=I'+r_xX'+r_yY'+r_zZ',
\end{equation}
are initially equal to 
\begin{eqnarray}
I'&=&I^{[1]}I^{[2]}-Z^{[1]}Z^{[2]},\\ 
X'&=&X^{[1]}X^{[2]}+Y^{[1]}Y^{[2]},\\
Y'&=&Y^{[1]}X^{[2]}-X^{[1]}Y^{[2]},\\
Z'&=&\frac{1}2(Z^{[1]}-Z^{[2]}). 
\end{eqnarray}
It is easy to verify that these operators satisfy the Pauli algebra relations, $[X',Y']=iZ'$, and cyclic permutations thereof, and $X'^2=Y'^2=Z'^2=I'$. 

We will now present analogs of Eqns. (\ref{ich15}-\ref{ich3}) for these operators. This can be done by expanding the solution of the Schr\"odinger equation into the Taylor series,
\begin{equation}
\label{taylor}
O(t)=O+\frac{i}{1!}[H,O]t+\frac{i^2}{2!}[H,[H,O]]t^2+...\quad .
\end{equation} 
For the moment let us take general inter-spin couplings $J_i$, rather than those from Eqn. (\ref{Hxx}). The first few commutators read
\begin{widetext}
\begin{eqnarray}
\label{twoqb}
[H,X^{[1]}X^{[2]}+Y^{[1]}Y^{[2]}]&=&iJ_2(-X^{[1]}Z^{[2]}Y^{[3]}+Y^{[1]}Z^{[2]}X^{[3]}),\\
\![H,[H,X^{[1]}X^{[2]}+Y^{[1]}Y^{[2]}]]&=&\frac{1}{4}J^2_2(X^{[1]}X^{[2]}+Y^{[1]}Y^{[2]})
-\frac{1}{4}J_1J_2(X^{[2]}X^{[3]}+Y^{[2]}Y^{[3]})\nonumber\\
&+&J_2J_3(X^{[1]}Z^{[2]}Z^{[3]}X^{[4]}+Y^{[1]}Z^{[2]}Z^{[3]}Y^{[4]})\\
&&...\quad,\nonumber\\
\![H,Y^{[1]}X^{[2]}-X^{[1]}Y^{[2]}]&=&i\frac{1}{2}J_1(Z^{[1]}-Z^{[2]})-iJ_2(Y^{[1]}Z^{[2]}Y^{[3]}+X^{[1]}Z^{[2]}X^{[3]}),\\
\![H,[H,Y^{[1]}X^{[2]}-X^{[1]}Y^{[2]}]]&=&-\frac{3}{4}J_1J_2(Y^{[2]}X^{[3]}-X^{[2]}Y^{[3]})+(J_1^2+\frac{1}{4}J_2^2)(Y^{[1]}X^{[2]}-X^{[1]}Y^{[2]})\nonumber\\
&+&J_2J_3(Y^{[1]}Z^{[2]}Z^{[3]}X^{[4]}-X^{[1]}Z^{[2]}Z^{[3]}Y^{[4]}),\\
&&...\quad,\nonumber\\
\![H,Z^{[1]}-Z^{[2]}]&=&2iJ_1(X^{[1]}Y^{[2]}-X^{[1]}Y^{[2]})+iJ_2(Y^{[2]}X^{[3]}-X^{[2]}Y^{[3]}),\\ 
\![H,[H,Z^{[1]}-Z^{[2]}]]&=&J_1^2(Z^{[1]}-Z^{[2]})-3J_1J_2(X^{[1]}Z^{[2]}X^{[3]}+Y^{[1]}Z^{[2]}Y^{[3]})\nonumber\\
&&+\frac{1}{2}J_2^2(Z^{[3]}-Z^{[2]})+J_2J_3(X^{[2]}Z^{[3]}X^{[4]}+Y^{[2]}Z^{[3]}Y^{[4]}),\\
&&...\quad,\nonumber\\
\![H,Z^{[1]}Z^{[2]}]&=&iJ_2(Z^{[1]}X^{[2]}Y^{[3]}-Z^{[1]}Y^{[2]}X^{[3]}),\\
\![H,[H,Z^{[1]}Z^{[2]}]]&=&\frac{1}{4}J_1J_2(X^{[1]}X^{[3]}+Y^{[1]}Y^{[3]})+\frac{1}{2}J_2^2(Z^{[1]}Z^{[2]}-Z^{[1]}Z^{[3]})\nonumber\\
\label{twoqe}
&&-J_2J_3(Z^{[1]}X^{[2]}Z^{[3]}X^{[4]}+Z^{[1]}Y^{[2]}Z^{[3]}Y^{[4]}),\\
&&...\quad.\nonumber
\end{eqnarray}
\end{widetext}

Every second iteration of the commutator produces the term which is similar to the original operator, but shifted by one position toward the other end of the chain. Only in case of $Z^{[1]}Z^{[2]}$ does it take four iterations to get the same effect. 

From the property of state mirroring it follows that at time $t^*$ all terms with operators acting on qubits different than the last two vanish. At this instance, the analogues of Eqns. (\ref{ich15}-\ref{ich3}) read
\begin{eqnarray}
I(t^*)&=&I^{[N-1]}I^{[N]}-Z^{[N-1]}Z^{[N]},\\
\label{nasze15}
X'(t^*)&=&X^{[N-1]}X^{[N]}+Y^{[N-1]}Y^{[N]},\\
\label{nasze23}
Y'(t^*)&=&X^{[N-1]}Y^{[N]}-Y^{[N-1]}X^{[N]},\\
Z'(t^*)&=&\frac{1}{2}(Z^{[N-1]}-Z^{[N]}).
\end{eqnarray}
Hence, the full transfer is described by 
\begin{eqnarray}
\label{fulltransfer}
&\rho^{[1,2]}\otimes\sigma^{[3,4,...,N]}\stackrel{t^*}{\longrightarrow}&\nonumber\\&\sigma'^{[N-2,N-3,...,1]}\otimes\rho^{[N,N-1]}&
\end{eqnarray}
with $\text{Tr}\rho^{[1,2]}Z^{[1]}Z^{[2]}=-\frac{1}{4}$ and $\sigma$ is an arbitrary $(N-2)$-qubit state, which is, in general, changed during the transfer. The receiver has full access to the quantum message, independently of the state of the rest of the wire. 

This is in contrast to the usual one qubit encoding, where the evolution generates $N$-spin operators as follows:
\begin{eqnarray}
\label{oneqb}
\![H,X^{[1]}]&=&iJ_1Z^{[1]}Y^{[2]},\\
\![H,[H,X^{[1]}]]&=&\frac{1}{4}J_1^2X^{[1]}+J_1J_2Z^{[1]}Z^{[2]}X^{[3]},\\
&...&\quad,\nonumber\\
\label{oneqe}
\![H,Y^{[1]}]&=&-iJ_1Z^{[1]}X^{[2]},\\
\![H,[H,Y^{[1]}]]&=&\frac{1}{4}J_1^2Y^{[1]}+J_1J_2Z^{[1]}Z^{[2]}Y^{[3]},\\
&...&\quad.\nonumber
\end{eqnarray}

The above equations allow better understanding of the phenomenon of the state transfer without state initialization and remote collaboration. Consider the transfer of state $(|0\rangle+|1\rangle)/\sqrt{2}$, with the one qubit encoding. In the standard way, the initial state of the whole chain is
\begin{eqnarray}
\label{onestart}
&\frac{1}{2}\left((|0\rangle+|1\rangle)\otimes|00...\rangle\right)\left((\langle 0|+\langle 1|)\otimes\langle 00...|\right)&\nonumber\\
=&(I^{[1]}+X^{[1]})(I^{[2]}+Z^{[2]})(I^{[3]}+Z^{[3]})...\!,
\end{eqnarray}
whereas in the two spin encoding we have
\begin{eqnarray}
\label{twostart}
&\frac{1}{2}\left((|10\rangle+|01\rangle)\otimes|00..\rangle\right)\left((\langle 10|+\langle 01|)\otimes\langle 00...|\right)&\nonumber\\
=&(I^{[1]}I^{[2]}+X^{[1]}X^{[2]}+Y^{[1]}Y^{[2]}-Z^{[1]}Z^{[2]})\nonumber\\
\times&(I^{[3]}+Z^{[3]})(I^{[4]}+Z^{[4]})... \!.
\end{eqnarray}

As the system is state mirroring, both superpositions are perfectly reconstructed at the other end of the chain at time $t^*$. However, if one applies Eqns. (\ref{oneqb}-\ref{oneqe}) and others of this kind, one sees that $X^{[1]}$ evolved to $...Z^{[N-2]}Z^{[N-1]}X^{[N]}$ (or $...Z^{[N-2]}Z^{[N-1]}Y^{[N]}$, depending on the parity of $N$), $X^{[1]}Z^{[2]}$ to $...Z^{[N-2]}X^{[N]}$ ($...Z^{[N-2]}Y^{[N]}$), and so on. In particular, it was the evolution $X^{[1]}Z^{[2]}...Z^{[N]}$ that generated the $X^{[N]}$ ($Y^{[N]}$). It is explicit that with the traditional encoding the transfer relies on correlations of the rest of the state. 

In conclusions, we have presented a method to communicate a qubit over a chain of spins (quantum wire) without a need for the global state initialization. Our result has significant advantages over the one proposed in \cite{paternostro}. We do not require any additional communication, conditional transformation, or, most importantly, simultaneous state manipulation by both users. It is worth stressing, that in the presented protocol the state is initialized only in the sender's site. Another advantage is that the transfer could be signaled to the receiver after the encoding. The speed of classical message is expected to be much higher than that of quantum information. The only additional requirement in our protocol is that the observers have the access to two, not one end spins. This allows find of a two-dimensional code subspace, which is mirrored in a quantum wire independently of the state of the rest of the chain. In a natural way, our result applies not only to the particular system discussed in this manuscript, but equally well to all systems with the state mirroring property, e. g., the families discussed in \cite{Shi}. 

It should be emphasized, that the transfer protocol might be an element of a very composite computational process. We propose a significant simplification of the transfer routine without initialization of the state of the interconnecting part of the chain. The reduction in the complexity of elementary tasks contributes to the simplification of the whole computation. 

Moreover, the presented way of transferring a qubit possesses greater robustness against some specific noise models \cite{wiesniak}. Therein, it was shown that this code provides higher average fidelity when the wire weakly interacts with a global thermal bath at high temperatures.

The work is part of EU 6FP programmes QAP and SCALA and has been done at the National
Centre for Quantum Information at Gda\'nsk.


\begin{thebibliography}{99}
\bibitem{DJ} D. Deutsch and R. Jozsa, {\em Proc. R. Soc. London, Ser. A} 439, 553 (1992).
\bibitem{Shor} P. W. Shor, {\em Proceedings of the 35th Annual Symposium on Foundations of Computer Science, Los Alamos} (IEEE Computer Society Press, Los Alamitos, CA, 1994).
\bibitem{Grover} L. Grover, {\em Proceedings of the 28th Annual ACM Symposium on the Theory of Computing} (SOTC) (ACM Press, New York, 1996), p. 212.
\bibitem{jj1} J.Q. You, J.S. Tsai, F. Nori {\em Phys. Rev. Lett.} {\bf 89}, 197902 (2002).
\bibitem{jj2} A. Lyakhov and C. Bruder, {\em New Jour. Phys.} {\bf 7}, 181, (2005).
\bibitem{jj3} A. Romito, R. Fazio, and C. Bruder, {\em Phys. Rev. B} {\bf 71}, 100501 (2005).
\bibitem{qd1} D. Loss and D. P. DiVincenzo, {\em Phys. Rev. A} {\bf 57}, 120 (1998).
\bibitem{qd2} B. E. Kane, {\em Nature} (London) {\bf 393}, 133 (1998).
\bibitem{qd3} R. Vrijen, E. Yablonovitch, K.Wang, H.W. Jiang, A. Balandin, V. Roychowdhury, T. Mor, and D. DiVincenzo, 
{\em Phys. Rev. A} {\bf 62}, 012306 (2000).
\bibitem{sugato} S. Bose, {\em Phys. Rev. Lett.} {\bf 91}, 207901 (2003).
\bibitem{bgb} D. Burgarth, V. Giovannetti, and S. Bose, {\em Phys. Rev. A} {\bf 75},062327 (2007).
\bibitem{danielbose} D. Burgarth and S. Bose, {\em Phys. Rev. A} {\bf 71}, 052315 (2005).
\bibitem{danielbose2} D. Burgarth and S. Bose, {\em New Jour. Phys.} {\bf 7}, 135 (2005).
\bibitem{poznan} A. W\'ojcik, T. \L uczak, P. Kurzy\'nski, A. Gr\'odka, T. Gdala, and M. Bednarska, {\em Phys. Rev. A}{\bf 72}, 034303 (2005). 
\bibitem{christ} M. Christandl, N. Datta, A. Ekert, and A. J. Landahl, {\em Phys.
Rev. Lett.} {\bf 92}, 187902 (2004).
\bibitem{niko1} G. M. Nikolopoulos, D. Petrosyan, and P. Lambropoulos, {\em Europhys.
Lett.} {\bf 65}, 297 (2004).
\bibitem{niko2} G. M. Nikolopoulos, D. Petrosyan, and P. Lambropoulos, {\em J.
Phys.: Condens. Matter} {\bf 16}, 4991 (2004).
\bibitem{Shi} T. Shi, Y. Li, Z. Song, and C. P. Sun, Phys. Rev. A 71, 032309 (2005).
\bibitem{danielkoji} D. Burgarth, K. Maruyama, and F. Nori, {\em Phys. Rev. A} {\bf 79} 020305(R) (2009).
\bibitem{paternostro2} C. Di Franco, M. Paternostro, M. S. Kim, e-print arXiv:0812.3510; {\em Phys. Rev. Lett.} (to be published).
\bibitem{paternostro} C. Di Franco, M. Paternostro, and M. S. Kim, {\em Phys. Rev. Lett.} {\bf 101}, 230502 (2008).
\bibitem{teleportation} C. H. Bennett, G. Brassard, S. Popescu, B. Schumacher,
J.A. Smolin, and W. K. Wootters, {\em Phys. Rev. Lett.} {\bf 76},
722 (1996).
\bibitem{wiesniak} M. Wie\'sniak, arxiv:0711.2357; M. Markiewicz and M. Wie\'sniak, in preparation.
\end{thebibliography}
\end{document}